%% file: resparsification.tex
\documentclass[11pt,letterpaper]{article}

\input{macros.tex}
\input{bib_macros.tex}

\begin{document}

\title{Analysis of Resparsification\\{(draft)}}
\author{Jakub Pachocki\\\texttt{pachocki@cs.cmu.edu}}
\date{}
\maketitle

\begin{abstract}
    We show that schemes for sparsifying matrices based on iteratively resampling rows yield guarantees matching classic 'offline' sparsifiers (see e.g. \cite{SpielmanS08}).
    In particular, this gives a formal analysis of a scheme very similar to the one proposed by Kelner and Levin \cite{KelnerL13}.
\end{abstract}

\section{Introduction}
In this note, we deal with the problem of spectrally approximating, or \emph{sparsifying}, a tall $n \times d$ matrix $\bv{A}$.
Our goal will be to find an $\Oh(d \log d \epsilon^{-2})\times d$ matrix $\bv{\tilde{A}}$ such that $\norm{\bv{\tilde{A}}\bv{x}} \approx \norm{\bv{A}\bv{x}}$ for all $\bv{x}$; we will focus on the streaming setting, where we read rows of $\bv{A}$ one by one.
We will also require that our $\bv{\tilde{A}}$ preserves the structure of $\bv{A}$: that is, it will only consist of a subset of (reweighted) rows of $\bv{A}$.

It is well known that sampling $\Oh(d \log d \epsilon^{-2})$ rows of $\bv{A}$ with probabilities proportional to their \emph{leverage scores} yields a $(1+\epsilon)$-spectral approximation to $\bv{A}$ (see e.g. \cite{SpielmanS08, LiMP13, CohenLMMPS15, CohenMP16}).

Kelner and Levin \cite{KelnerL13} proposed a scheme for computing $\bv{\tilde{A}}$ through simply storing the rows of $\bv{A}$ in memory as they are read, and applying the above sparsification procedure whenever the matrix becomes too large.
The main contribution of this note is a rigorous proof of such a scheme.

We believe the fact that the \emph{resparsification} paradigm yields algorithms with guarantees as good as offline sparsifiers has implications reaching beyond the streaming setting; similar schemes have been used for example in designing fast Laplacian solvers \cite{PengS14,KyngS16}.

\section{Preliminaries}

Let $\bv{a}_1,\ldots,\bv{a}_n$ be vectors in $\mathbb{R}^d$, such that
\begin{align*}
    \sum_{i=1}^n \bv{a}_i \bv{a}_i^\top = \bv{I}.
\end{align*}

We will use the following two theorems due to Tropp \cite{Tropp11,Tropp12}:
\begin{theorem}[Matrix Chernoff]
    \label{thm:chernoff}
    Consider a finite sequence $\{\bv{X}_k\}$ of independent, random, self-adjoint matrices with dimension $d$.
    Assume that each random matrix satisfies
    \begin{align*}
        \bv{X}_k \succeq \bv{0} \mbox{ and } \norm{\bv{X}_k} \leq R \mbox{ almost surely.}
    \end{align*}
    Define
    \begin{align*}
        \mu_{\max} := \norm{\sum_k \expct{\bv{X}_k}}.
    \end{align*}
    Then
    \begin{align*}
        \prob{\norm{\sum_k \bv{X}_k} \geq (1+\delta)\mu_{\max}} \leq d\cdot \left(\frac{e^\delta}{(1+\delta)^{1+\delta}}\right)^{\mu_{\max}/R} \mbox{ for $\delta \geq 0$.}
    \end{align*}
\end{theorem}

\begin{theorem}[Matrix Freedman]
    \label{thm:freedman}
    Let $\bv{Y}_0, \bv{Y}_1, \bv{Y}_2, \ldots$ be a matrix martingale whose values are self-adjoint matrices with dimension $d$, and let $\bv{X}_1, \bv{X}_2, \ldots$ be the difference sequence.
    Assume that the difference sequence is uniformly bounded in the sense that
    \begin{align*}
        \norm{\bv{X}_k} &\leq R\mbox{ almost surely, for all } k.
    \end{align*}
    Define the predictable quadratic variation process of the martingale:
    \begin{align*}
        \bv{W}_k := \sum_{j=1}^k \expctt{j - 1}{\bv{X}_j^2}\mbox{, for all } k.
    \end{align*}
    Then, for all $t > 0$ and $\sigma^2 > 0$,
    \begin{align*}
        \prob{\exists k : \norm{\bv{Y}_k} \geq t \mbox{ and } \norm{\bv{W}_k} \leq \sigma^2} &\leq d\cdot\exp\left(-\frac{-t^2/2}{\sigma^2+Rt/3}\right).
    \end{align*}
\end{theorem}

\section{The Game}

Fix $\epsilon \in \left(0, \frac{1}{2}\right)$.
Set $c = 100 \epsilon^{-2} \log d$.
Assume that for all $i$, $\bv{a}_i^\top \bv{a}_i \leq \frac{1}{c}$ (this assumption is technical and not actually necessary; larger rows are simply ignored in the game anyway).
For $i = 1,\ldots,n$, set $w_i$ to $1$.

We will analyze a game played by an adversary on the weights $w_i$.
It consists of a single move, repeated while the game is not over:

\begin{enumerate}
    \item The adversary picks any $i$ such that $w_i \neq 0$ and $2 w_i \bv{a}_i^\top \bv{a}_i \leq \frac{1}{c}$.
    \item Flip an unbiased coin.
        If it comes out heads, set $w_i \leftarrow 2w_i$; otherwise, set $w_i \leftarrow 0$.
\end{enumerate}

The game can end in one of two ways:
\begin{itemize}
    \item The matrix $\sum w_i \bv{a}_i \bv{a}_i^\top$ is not a $(1 + \epsilon)$-approximation to the identity; then, the adversary wins.
    \item The adversary has no more legal moves; then, the adversary loses.
\end{itemize}

We will show that with high probability, the adversary will not win the game.
\begin{theorem}
\label{thm:main}
    With high probability, the game defined above ends in a loss for the adversary.
\end{theorem}

\section{Bounding the Total Variation}

For all $i$, let $w_i'$ be the maximum value attained by $w_i$ throughout the game.

\begin{lemma}
    \label{lem:gamebound}
    Whp, we have that
    \begin{align*}
        \norm{\sum w_i'^2 (\bv{a}_i \bv{a}_i^\top)^2} \leq \frac{4}{c}.
    \end{align*}
\end{lemma}
\Proof
First of all, we consider a slightly modified version of the game: if the game ends in a victory for the adversary, we still let them perform legal moves while possible.
This can only increase the $w_i'$s.

Now note that in this augmented game, the adversary's choices do not have any effect on the final values of $w_i$ and $w_i'$s; for every $i$, the adversary will keep increasing $w_i$ until it is either too large or $0$.
The $w_i'$ are therefore independent.

Let $\bv{X}_i' := w_i'^2 (\bv{a}_i\bv{a}_i^\top)^2$, for $i = 1, \ldots, n$.
Our goal is now to bound the norm of the sum of the independent matrices $\bv{X}_i'$.
First of all, note that we always have $\norm{\bv{X}_i'} \leq \frac{1}{c^2}$.

Let $B_i$ be the maximum integer such that $2^{B_i} \bv{a}_i^\top \bv{a}_i \leq \frac{1}{c}$.
We have

\begin{align*}
    \expct{\bv{X}_i'}  &\preceq \sum_{k=0}^{B_i} 2^{-k} \cdot 4^k \cdot (\bv{a}_i\bv{a}_i^\top)^2\\
                       &\preceq 2^{B_i + 1} \cdot (\bv{a}_i\bv{a}_i^\top)^2\\
                       &\preceq \frac{2}{c} \cdot \bv{a}_i\bv{a}_i^\top,\\
\end{align*}

and so
\begin{align*}
    \norm{\sum_{i=1}^n \expct{\bv{X}_i'}} \leq \frac{2}{c}.
\end{align*}

The thesis follows from \Cref{thm:chernoff} with $\delta = \frac{2}{c\cdot\mu_{\max}}, R = \frac{1}{c^2}$.
\QED

\section{The Proof}

We consider a martingale with the difference $\bv{X}_j$ corresponding to the $j$-th move by the adversary, or $\bv{0}$ if the adversary made less than $j$ moves.
Assume the adversary chooses row $\bv{a}_i$; then we have
\begin{align*}
    \bv{X}_j := 
    \begin{cases}
        w_i \bv{a}_i\bv{a}_i^\top &\mbox{ with probability $\frac{1}{2}$} \vspace{0.2cm}\\
        -w_i \bv{a}_{i}\bv{a}_i^\top &\mbox{ otherwise.}\\
    \end{cases}
\end{align*}

Let $\{\bv{W}_k\}$ be the predictable quadratic variation process of the martingale given by the $\{\bv{X}_j\}$:
\begin{align*}
    \bv{W}_k := \sum_{j=1}^k \expctt{j-1}{\bv{X}_j^2}.
\end{align*}

\begin{lemma}
\label{lem:loww}
    Whp, for all $k$ we have that
    \begin{align*}
        \norm{\bv{W}_k} \leq \frac{8}{c}.
    \end{align*}
\end{lemma}
\Proof
We have
\begin{align*}
    \expctt{j-1}{\bv{X}_j^2} &= w_i^2 (\bv{a}_{i}\bv{a}_i^\top)^2.\\
\end{align*}
Therefore
\begin{align*}
    \bv{W}_k &\preceq \sum_{i=1}^n 2 \cdot w_i'^2 (\bv{a}_i\bv{a}_i^\top)^2.
\end{align*}
The thesis follows from \Cref{lem:gamebound}.
\QED

\Proofof{\Cref{thm:main}}
Note that $\norm{\bv{X}_k} \leq \frac{1}{c}$.
Together with \Cref{lem:loww}, applying \Cref{thm:freedman} with $t = \epsilon, \sigma^2 = \frac{8}{c}, R = \frac{1}{c}$ yields the thesis.
\QED

\section{Application to Streaming Sparsification}
\label{sec:application}

Consider any sparsification algorithm that reads edges of the graph one by one and adds them to the sparsifier.
At any time, the algorithm is free to take any edges whose leverage score is not too large, and remove them from the sparsifier with probability $\frac{1}{2}$ and double their weight otherwise.
Such an algorithm can be implemented in $\Oh(d \log d \epsilon^{-2})$ space and nearly linear time (see \Cref{fig:streaming-sample}).

\Cref{thm:main} gives that any such algorithm will output a sparsifier of the original graph whp. at the end.
Some additional care is required to show that we can maintain a sparsifier to the current graph at all times.

\begin{figure}[ht]
\noindent
\centering
\fbox{
\begin{minipage}{6in}
    \noindent $\bv{\tilde{A}} = \textsc{Streaming-Sample} (\bv{A}, \epsilon)$,
    where $\bv{A}$ is an $n \times d$ matrix with rows $\bv{a}_1, \ldots, \bv{a}_n$, $\epsilon \in (0, \frac{1}{2})$.
\begin{enumerate}
\item Set $c := 100 \log d / \epsilon^2$.
\item Let $\bv{\tilde{A}}$ be a $0 \times d$ matrix.
\item For $i = 1, \ldots, n$:
    \begin{enumerate}
        \item Append $\bv{a}_i$ to $\bv{\tilde{A}}$.

        \item If $\bv{\tilde{A}}$ has more than $20 d c$ rows:
            \begin{enumerate}
                \item\label{step:lev} Let $l_j$ be the leverage score of the $j$-th row of $\bv{\tilde{A}}$.
                \item While $\bv{\tilde{A}}$ has more than $10 d c$ rows:
                \begin{enumerate}
                    \item Pick an arbitrary row $\bv{\tilde{a}}_j$ of $\bv{\tilde{A}}$ with $l_j$ less than $\frac{1}{4c}$.
                    \item With probability $\frac{1}{2}$, remove $\bv{\tilde{a}}_j$ from $\bv{\tilde{A}}$; otherwise, double $\bv{\tilde{a}}_j$ and $l_j$.
                \end{enumerate}
            \end{enumerate}
    \end{enumerate}
\item Return $\bv{\tilde{A}}$.
\end{enumerate}
\end{minipage}
}
\caption{The resparsifying streaming algorithm.}
\label{fig:streaming-sample}
\end{figure}

\begin{theorem}
    Let $\bv{\tilde{A}}$ be the matrix returned by $\textsc{Streaming-Sample}(\bv{A}, \epsilon)$. Then, with high probability,
    \begin{align*}
        (1-\epsilon)\bv{A}^\top\bv{A} \preceq \bv{\tilde{A}}^\top\bv{\tilde{A}} \preceq (1+\epsilon)\bv{A}^\top\bv{A}.
    \end{align*}
\end{theorem}
\Proofsketch
To apply \Cref{thm:main}, it is enough to observe that, unless the algorithm already failed, the value $l_j$ computed in step~\ref{step:lev} of $\textsc{Streaming-Sample}$ is at most $1+\epsilon$ times smaller than the leverage score of $\bv{\tilde{a}}_j$ wrt. $\bv{A}$.
\QED

\section{Acknowledgments}
The author would like to thank Richard Peng for suggesting the usage of sampling through unbiased coin flips, which significantly simplified the analysis.
The author would also like to thank Michael Cohen and Yiannis Koutis for helpful comments and feedback.

\bibliographystyle{alpha}
\bibliography{resparsification}

\end{document}

%% file: macros.tex
\usepackage{thmtools}
\usepackage{thm-restate}

\usepackage{mathrsfs}           
\usepackage{vmargin,fancyhdr}   
\usepackage{enumerate}
\usepackage{enumitem}

\usepackage{amsmath,amssymb}    
\usepackage{verbatim}           
\usepackage{xspace}             
\usepackage{graphicx,float}     
\usepackage{ifthen,calc}        
\usepackage{textcomp}           
\usepackage{fancybox}           
\usepackage{hhline}             
\usepackage{float}              

\usepackage[rflt]{floatflt}     
\usepackage[small,compact]{titlesec}
\usepackage{setspace}
\usepackage{subfigure}

\usepackage{color}
\definecolor{ForestGreen}{rgb}{0.1333,0.5451,0.1333}
\usepackage{thumbpdf}
\usepackage[letterpaper,
            colorlinks,linkcolor=ForestGreen,citecolor=ForestGreen,
            backref,
            bookmarks,bookmarksopen,bookmarksnumbered]
           {hyperref}

\usepackage{cleveref}

\setpapersize{USletter}
\setmarginsrb{1in}{.5in}        
             {1in}{.5in}        
             {.25in}{.25in}     
             {.25in}{.5in}      
\newcommand{\showccc}[0]{0}
\newcommand{\ccc}[2][nothing]{
  \ifthenelse{\showccc=0}{}{
    \ensuremath{^{\Lsh\Rsh}}\marginpar{\raggedright\tiny\textsf{%
        \ifthenelse{\equal{#1}{nothing}}{}{\textbf{#1}\\}#2}}}}

\newcounter{hours}\newcounter{minutes}
\newcommand{\hhmm}{%
  \setcounter{hours}{\time/60}%
  \setcounter{minutes}{\time-\value{hours}*60}%
  \ifthenelse{\value{hours}<10}{0}{}\thehours:%
  \ifthenelse{\value{minutes}<10}{0}{}\theminutes}
\ifthenelse{\showccc=0}{\rhead{}}{\rhead{\today \ [\hhmm]}}
\newtheorem{theorem}{Theorem}[section]

\newtheorem{lemma}[theorem]{Lemma}

\newcommand{\Proof}[0]{\smallskip\noindent\textit{\textbf{Proof:}}\quad}
\newcommand{\Proofsketch}[0]{\smallskip\noindent\textit{\textbf{Sketch of proof:}}\quad}
\newcommand{\Proofof}[1]{\smallskip\noindent\textit{\textbf{Proof of #1:}}\quad}

\newcommand{\Oh}{\ensuremath{\mathcal{O}}}

\newcommand{\bv}[1]{\mathbf{#1}}

\newcommand{\QED}[0]{\hfill\ensuremath{\blacksquare}\medspace\\}

\floatstyle{ruled}
\newfloat{algo}{tbp}{lop}
\floatname{algo}{Algorithm}

\usepackage{float}
\floatstyle{plain}\newfloat{myfig}{t}{figs}[section]
\floatname{myfig}{\textsc{Figure}}
\floatstyle{plain}\newfloat{myalg}{H}{algs}[section]
\floatname{myalg}{}
\newcommand{\norm}[1]{\left\|#1\right\|_2}

\newcommand{\expct}[1]{\ensuremath{\mathbb{E}\left[#1\right]}}
\newcommand{\expctt}[2]{\ensuremath{\mathbb{E}_{#1}\left[#2\right]}}
\newcommand{\prob}[1]{\ensuremath{\mathop{\text{\normalfont \textbf{P}}}}\left[#1\right]}

\usepackage{tikz}
\usetikzlibrary{positioning,chains,fit,shapes,calc}

%% file: bib_macros.tex
  \usepackage{nth}
  \usepackage{intcalc}